\documentclass[review]{elsarticle}
\usepackage{lineno,hyperref}
\usepackage{graphicx}
\usepackage{float} 
\usepackage{subfigure}
\modulolinenumbers[5]
\usepackage{epsfig}
\usepackage{makeidx}
\usepackage{natbib}
\usepackage{multirow}
\usepackage{multicol}
\usepackage{xcolor}
\usepackage{bm}
\usepackage{ulem}
\begin{document}

\begin{frontmatter}

\title{Two-stage multi-scale breast mass segmentation for full mammogram analysis without user intervention}

\author[address1,address2]{Yutong Yan}
\author[address2,address3]{Pierre-Henri Conze\corref{mycorrespondingauthor}}
\cortext[mycorrespondingauthor]{Corresponding author}
\ead{pierre-henri.conze@imt-atlantique.fr}
\author[address2]{Gwenol\'e Quellec}
\author[address1,address2]{Mathieu Lamard}
\author[address1,address2,address4]{Beatrice Cochener}
\author[address2,address3]{Gouenou Coatrieux}
\address[address1]{Universit\'e de Bretagne Occidentale, Brest, France}
\address[address2]{Inserm, LaTIM UMR 1101, Brest, France}
\address[address3]{IMT Atlantique, Brest, France}
\address[address4]{University Hospital of Brest, Brest, France} 

\begin{abstract}

\noindent Mammography is the primary imaging modality used for early detection and diagnosis of breast cancer. X-ray mammogram analysis mainly refers to the localization of suspicious regions of interest followed by segmentation, towards further lesion classification into benign versus malignant. Among diverse types of breast abnormalities, masses are the most important clinical findings of breast carcinomas. However, manually segmenting breast masses from native mammograms is time-consuming and error-prone. Therefore, an integrated computer-aided diagnosis system is required to assist clinicians for automatic and precise breast mass delineation. In this work, we present a two-stage multi-scale pipeline that provides accurate mass contours from high-resolution full mammograms. First, we propose an extended deep detector integrating a multi-scale fusion strategy for automated mass localization. Second, a convolutional encoder-decoder network using nested and dense skip connections is employed to fine-delineate candidate masses. Unlike most previous studies based on segmentation from regions, our framework handles mass segmentation from native full mammograms without any user intervention. Trained on INbreast and DDSM-CBIS public datasets, the pipeline achieves an overall average Dice of 80.44\% on INbreast test images, outperforming state-of-the-art. Our system shows promising accuracy as an automatic full-image mass segmentation system. Extensive experiments reveals robustness against the diversity of size, shape and appearance of breast masses, towards better interaction-free computer-aided diagnosis.

\end{abstract}


\end{frontmatter}


\section{Introduction}
\label{sec:sec1}

\noindent Breast cancer is ranked first among all cancers in terms of frequency, accounting for 25\% of cancer cases and 15\% of cancer-related deaths \cite{torre2017global}. It is also the leading cause of cancer death among women from 20 to 59 years old \cite{CS2018}. X-ray mammography is known as a key tool for cost-effective early detection of breast abnormalities and help women prevent and fight against breast cancer. 

Among diverse types of breast abnormalities (mass, calcification, asymmetry, distortion...), masses are the most important clinical symptoms of breast carcinomas. Texture, shape and margin characteristics of masses play a key role for further breast tissue analysis \cite{virmani2019effect}. Despite massive screening, many patients are given heavy treatments by mistake due to the lack of second reading \cite{myers2015benefits}. Computer-aided diagnosis (CAD) systems for mammogram interpretation have been designed to avoid time-consuming and tedious second opinions. Recent systems rely on deep learning methods for their ability to outperform traditional approaches without hand-crafted features. However, due to the requirements in clinical practice, some studies report that current CAD tools are inefficient and not automatic enough to significantly improve diagnosis guidance \cite{lehman2015diagnostic}.
 
Low signal-to-noise ratio and variability in mass shapes and contours make mass segmentation challenging from whole mammograms. Most existing CAD tools focus on segmentation from low-resolution mammograms or from manually extracted suspicious areas \cite{li2018improved, Singh2020}. Even if those solutions largely simplify the segmentation process, they come at the cost of overall robustness and applicability in routine. First, mass patches are less representative than the entire image. Second, accurate pre-selected mass regions are not available during screening.

In this work, we address breast mass segmentation from native full X-ray mammograms, one of the most essential and challenging task towards efficient automated mammogram analysis. Related works mainly rely on one-stage deep architectures \cite{RFB15a, Singh2020, yan2019embc}. However, we argue that the tasks of localizing mass areas from mammograms and extracting precise boundaries for each mass are naturally two tasks with contradictory focuses: context-level semantic information for the former, resolution-level details for the latter. Addressing both challenges into one single network may lead to a sub-optimal trade-off and thus hinder precise full mammogram delineations. Additionally, increasing the network depth which could cover more spatial context and extract higher level features cannot be done \textit{ad-infinitum} for memory and computational reasons. 

\begin{figure}[htbp] 
\centering
\includegraphics[width=12cm]{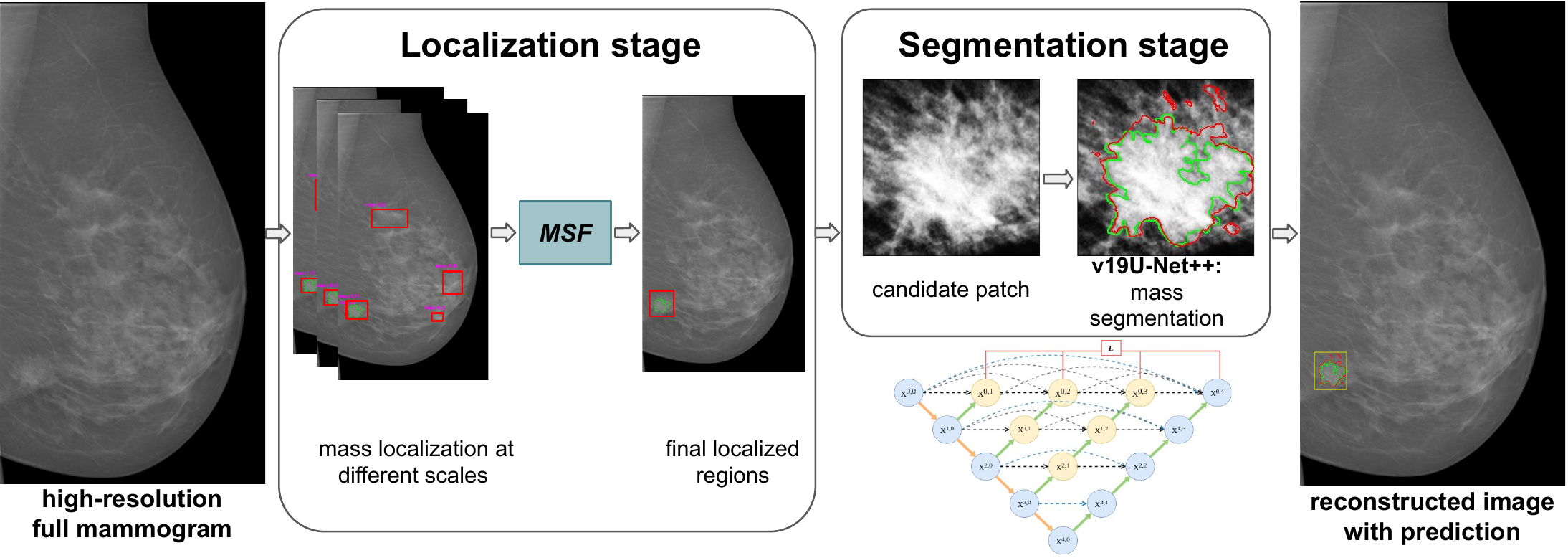}
\caption{Two-stage multi-scale pipeline for mass localization and segmentation from high-resolution X-ray mammograms. Red (green) lines indicate estimated (ground truth) delineations. MSF deals with the proposed multi-scale fusion strategy for automatic mass selection.}
\vspace{0.2cm}
\label{fig:Fig1}
\end{figure}

To tackle the aforementioned problems, we propose a two-stage multi-scale pipeline which performs automatic and highly precise mass segmentation from native high-resolution X-ray mammograms. The proposed framework (Fig.\ref{fig:Fig1}) consists of two steps. First, image-based mass localization using a novel multi-scale fusion is performed for automatic mass selection. Specifically, mass detector can be trained at multiple scales such that the localization procedure can also be extended by fusing predictions performed at multiple scales. Thus, we aim at identifying masses of any size, position or shape from the whole image, regardless of the resolution details. Second, we propose to employ a region-based mass segmentation model relying on a convolutional encoder-decoder architecture with nested and dense skip connections to obtain better mass delineations than standard deep segmentation models. Through our pipeline, we drastically reduce the number of unsuccessful detections while allowing a variable number of candidate regions to be automatically selected for segmentation without any expert interventions, leading to more reliable mass contours from full mammograms. The proposed approach can be easily integrated into clinical routine and is able to help diagnosis by acting as a relevant fully-automated second opinion.

This paper is organized as follows. In Sect.\ref{sec:sec2}, we present background material related to mass detection and segmentation using deep learning. The proposed two-stage framework associating mass localization and segmentation is presented in Sect.\ref{sec:sec3}. In particular, we describe a novel multi-scale fusion approach which improves the identification of suspicious areas. Sect.\ref{sec:sec4} provides experiments on public databases and prove the effectiveness of the whole framework. We end up with a discussion in Sect.\ref{sec:sec5} and conclusions in Sect.\ref{sec:sec6}.

\section{Related works}
\label{sec:sec2}

\noindent In the past few years, statistical models \cite{hizukuri2017segmentation} and machine learning techniques \cite{liu2020breast,hmida2017efficient} have been mainly used for lesion detection \cite{sapate2020breast}, classification \cite{liu2020breast,virmani2019effect,dalwinder2020simultaneous} and segmentation \cite{hizukuri2017segmentation,hmida2017efficient,oliver2015breast} tasks to assist clinicians for computer-assisted diagnosis of breast cancer. Some studies also focus on mammographic density characterization \cite{oliver2015breast,kanbayti2020mammographic,skarping2019mammographic} to target breast cancer management. In particular, Olivier et al. \cite{oliver2015breast} propose a pixel-based support vector machine (SVM) classifier for breast density segmentation. Hizukuri et al. \cite{hizukuri2017segmentation} introduce a level set method which is based on an energy function defined with region, edge and regularizing terms to segment breast masses. Hmida et al. \cite{hmida2017efficient} perform mass segmentation using a fuzzy active contour model obtained by combining fuzzy C-means and Chan-Vese models before classifying masses based on possibility theory. All these tasks are now routinely carried out in a purely data-driven fashion using convolutional neural networks (CNN). Specifically, many contributions have been proposed for breast imaging segmentation, which is an important and active research area. Deep learning-based segmentation is usually performed using convolutional encoder-decoder (CED) architectures such as fully convolutional networks (FCN) \cite{Long_2015_CVPR}, U-Net \cite{RFB15a} and Seg-Net \cite{badrinarayanan2015segnet} where the encoder performs multi-scale feature extraction whereas the symmetric decoder upsamples feature maps to recover spatial resolution. U-Net uses skip connections to combine corresponding encoder and decoder feature maps to better recover high-level details \cite{RFB15a} and works quite well with relatively small datasets.

The CED paradigm has been widely adopted by most of the recent approaches designed for breast mass segmentation. Owing to large but highly similar contextual features of mammograms and unpredictable shapes and sizes of masses, most segmentation techniques focus on pre-segmented regions of interest (RoI). Li et al. \cite{li2018improved} integrate the benefits of residual learning to improve the performance of U-Net to address gradient vanishing and exploding issues arising when increasing CNN depth. More recent studies introduce generative adversarial networks (GAN) \cite{GAN} where the adversarial network enforces the generative network to provide realistic contours. Thus, Singh et al. \cite{Singh2020} advocate conditional GAN with mass RoI as conditioning inputs for mass delineation. Caballo et al. \cite{caballo2020deep} also exploit GAN \cite{GAN} but as an augmentation strategy to generate synthetic breast images to further improve deep segmentation. Byra et al. \cite{byra2020breast} develop a selective kernel U-Net to adjust receptive fields through an attention mechanism and fuse feature maps with dilated and conventional convolutions. Alternatively, U-Net++ employs an CED with nested and dense skip connections \cite{zhou2018unet++}. However, these strategies focus on local segmentation while neglecting crucial contextual information. Afterwards, a multi-scale cascade of U-Net architectures as a one-stage full image mass segmentation method has been proposed \cite{yan2019embc} using multi-level image information fusion from high-resolution mammograms. Different from these works, we focus on a two-stage pipeline where masses are firstly localized before being precisely delineated.

Regarding breast mass detection, although many recently proposed object detection models \cite{YOLOv3,RenFasterRCNN,RFCN2016,GirshickDDM16} have achieved great success on common object detection tasks, automatic mass detection still remains a challenge due to the low signal-to-noise ratio and the unpredictable appearance of masses in X-ray mammograms. Sapate et al. \cite{sapate2020breast} propose a machine learning based algorithm to calculate the correspondence score of each lesion pair in dual views before fusing information to discriminate malignant tumours from benign masses using SVM. Agarwal et al. \cite{agarwal2019automatic} analyze the performance of popular deep CNN architectures in terms of mass/non-mass classification. Alternatively, Jung et al. \cite{Jung2018DetectionOM} propose a mass detector based on RetinaNet \cite{lin2017focal} exploiting a feature pyramid network optimized through a focal loss. Yap et al. \cite{yap2020breast} automate breast lesion detection using Faster-RCNN \cite{RenFasterRCNN} with Inception-ResNet-v2 \cite{szegedy2016inception}. However, these learning-based detectors may fail in identifying masses of any size, position or shape from the whole image. Existing detectors might therefore not produce sufficiently good proposals for further breast mass segmentation purposes.

Many studies focus on building multi-stage networks or integrating series of steps together. Dhungel et al. \cite{Dhungel2017ADL} propose a cascade of deep belief networks and Gaussian mixture models to provide mass candidates, followed by two cascades of CNN and random forest to refine detection results. Once suspicious areas are identified, they employ deep structured learning to perform mass segmentation. Al-Antari et al. \cite{Alantari2018AFI} propose an integrated mass detection, segmentation and classification pipeline from downsampled mammograms. Although their system could assist radiologists in multi-stage diagnosis, they still manually eliminate false localized candidate masses before the segmentation stage, which is impractical as an automatic CAD system. Apart from that, they exploit low-resolution mammograms. Image details are therefore lost during this process. In comparison, our approach aims at avoiding complex processing pipelines and human interventions, towards accurate and precise breast mass segmentation.

\section{Material and methods}
\label{sec:sec3}

\noindent To deal with mass segmentation from native resolution mammograms arising from public datasets such as INbreast \cite{Moreira2012INbreastTA} or DDSM-CBIS \cite{ddsm-cbis} (Sect.\ref{sec:sec3.1}), we propose an integrated framework (Fig.\ref{fig:Fig1}) consisting of two modules: image-based mass detection (Sect.\ref{sec:sec3.2}) followed by region-based mass segmentation (Sect.\ref{sec:sec3.3}). The former is based on a deep detection model extended based on a novel multi-scale fusion procedure to reduce wrong proposals and further improve detection accuracy (Sect.\ref{sec:sec3.2}). This stage performs coarse mass detection on entire mammograms and provides suspicious regions to the second stage. The latter conducts refined mass segmentation on extracted areas relying on a deep convolutional encoder-decoder architecture with nested and dense skip connections. An image reconstruction step is finally followed to visualize both mass location and segmentation results in high-resolution full mammograms.

\subsection{Imaging datasets}
\label{sec:sec3.1}

\noindent We focus on mass detection and segmentation from $2048 \times 1024$ full mammograms arising from two publicly-available mammogram datasets: INbreast \cite{Moreira2012INbreastTA} and DDSM-CBIS \cite{ddsm-cbis}. INbreast\footnotemark[1] \cite{Moreira2012INbreastTA} consists of 410 mammograms from 115 examinations. Four types of lesions (masses, calcifications, asymmetry and distortions) are included, but only 107 images containing masses as well as accurate delineations made by specialists are used. Conversely, DDSM-CBIS\footnotemark[2] (Digital Database for Screening Mammography) \cite{ddsm-cbis} is a relatively larger database containing approximately 2,500 mammograms including normal, benign and malignant cases and coarse ground truth manual delineations. In this work, 1514 DDSM-CBIS images containing masses are employed in the training phase.

\footnotetext[1]{\url{http://medicalresearch.inescporto.pt/breastresearch/}}
\footnotetext[2]{\url{https://doi.org/10.7937/K9/TCIA.2016.7O02S9CY}}

\subsection{Image-based mass detection}
\label{sec:sec3.2}

\noindent Among existing deep detectors, YOLOv3 \cite{YOLOv3} is adopted in this work for mass localization from full mammograms thanks to its good trade-off between accuracy and efficiency. However, other detectors such as SSD \cite{liu2015single}, Faster R-CNN \cite{RenFasterRCNN} or RetinaNet \cite{lin2017focal} can also be applied as alternative detection schemes. \\

\noindent \textbf{YOLOv3 detector.} The employed YOLO (You Only Look Once) implementation exploits the Darknet-53 backbone architecture consisting of 53 successive 3 $\times$ 3 and 1 $\times$ 1 convolutional layers as well as some shortcut connections. Feature maps from different scales are used to deal with huge mass size and aspect ratio variance, i.e., larger feature maps are assigned to detect smaller masses and vice versa. Following \cite{YOLOv3}, YOLOv3 uses anchor boxes to predict through regression the coordinates of bounding boxes. Different from Faster R-CNN \cite{RenFasterRCNN} which uses manually selected boxes, k-means clustering is used to recompute the 9 anchor settings to adapt YOLOv3 to the target mammography datasets. For training, we use pre-trained weights arising from ImageNet \cite{deng2009imagenet} pre-training. \\

\noindent \textbf{Extension using multi-scale fusion.} Although recently proposed detection models \cite{YOLOv3,RenFasterRCNN,RFCN2016,GirshickDDM16} have achieved excellent results on public common object detection datasets such as Pascal VOC \cite{Everingham15} or Microsoft-COCO \cite{LinMBHPRDZ14}, they are not optimal to be applied directly to mammograms for two main reasons. First, they are still struggling with object size variance. Typically, most object detectors have worse performance for small objects than for medium or large structures. Especially in our context, this problem becomes more serious as the size and aspect ratio of masses vary strongly. Second, mass detection is generally more difficult than common object detection since masses are visually less obvious and less contrasted with respect to surrounding healthy tissues, combined with a great diversity of shape and texture. Therefore, single-scale prediction might not provide sufficiently good proposals, leading to the failure of the next stage dedicated to breast mass segmentation.

\begin{figure}[htbp]
\centering
\includegraphics[width=12cm]{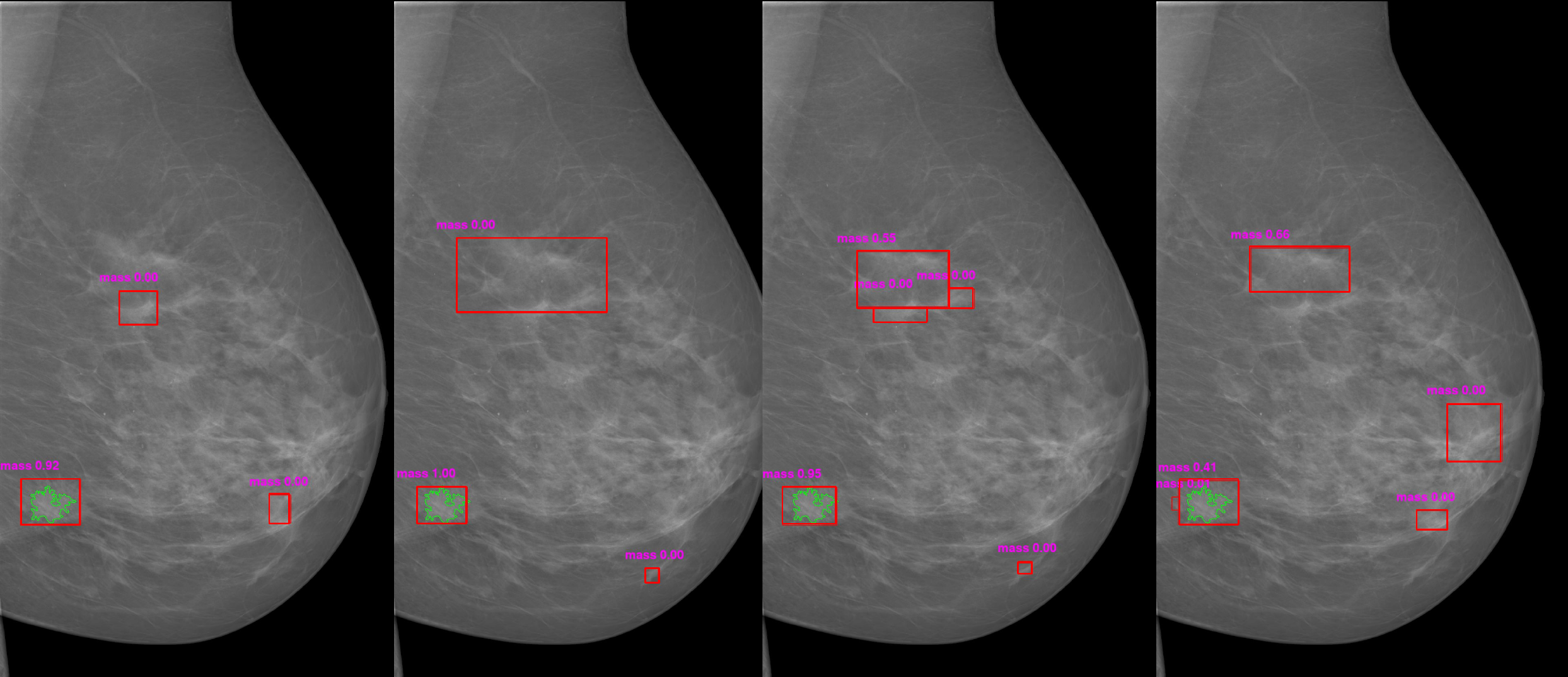} \vspace{-0.1cm} \\
\footnotesize \hspace{-0.5cm} $160 \times 320$ \hspace{1.6cm} $256 \times 512$ \hspace{1.5cm} $320 \times 640$\hspace{1.8cm} $480 \times 960$ \vspace{-0.1cm} \\
\caption{YOLOv3 predictions performed at multiple scales for one given mammogram. Red boxes correspond to mass RoI candidates with associated probabilities in magenta. Green contours arise from ground truth annotations.}
\label{fig:Fig2}
\end{figure}

In addition, previous works including \cite{Alantari2018AFI} that also use YOLO as mass detection model tend to manually select candidate masses to avoid false-positive proposals before the segmentation stage. We argue, however, that such approaches assume that they have already box-level expert annotations during validation and test phases, which is less practical and not obvious. As a matter of fact, an automatic and fully-integrated CAD system should not require any expert annotations for clinical purposes.

To address the problem of unsuccessful single-scale detection and avoid manual selection, we propose a multi-scale fusion (MSF) strategy. Note that one of the important designs in YOLOv3 is the multi-scale training, for which input images are dynamically resized every 10 batches instead of fixing the input image resolution. Image resolutions are randomly chosen from multiples of 32 since the model downsamples by a factor of 32. As a consequence, our MSF extension tends to fully exploit the multi-scale features extracted by YOLOv3 during training to further refine the generated candidates. Moreover, it allows us to be robust to the input size so that images with different resolutions can be processed without multiple training. In the same spirit as for training, we propose in the prediction stage to exploit results from different resolutions to make the network being more sensitive to masses with very small or large spatial extents. As shown in Fig.\ref{fig:Fig2}, we are able to perform different predictions at different scales using the same network. Thus, for a given mammogram, we propose to fuse predictions arising from multiple scales.

\begin{figure}[t]
\centering
{\epsfig{file = 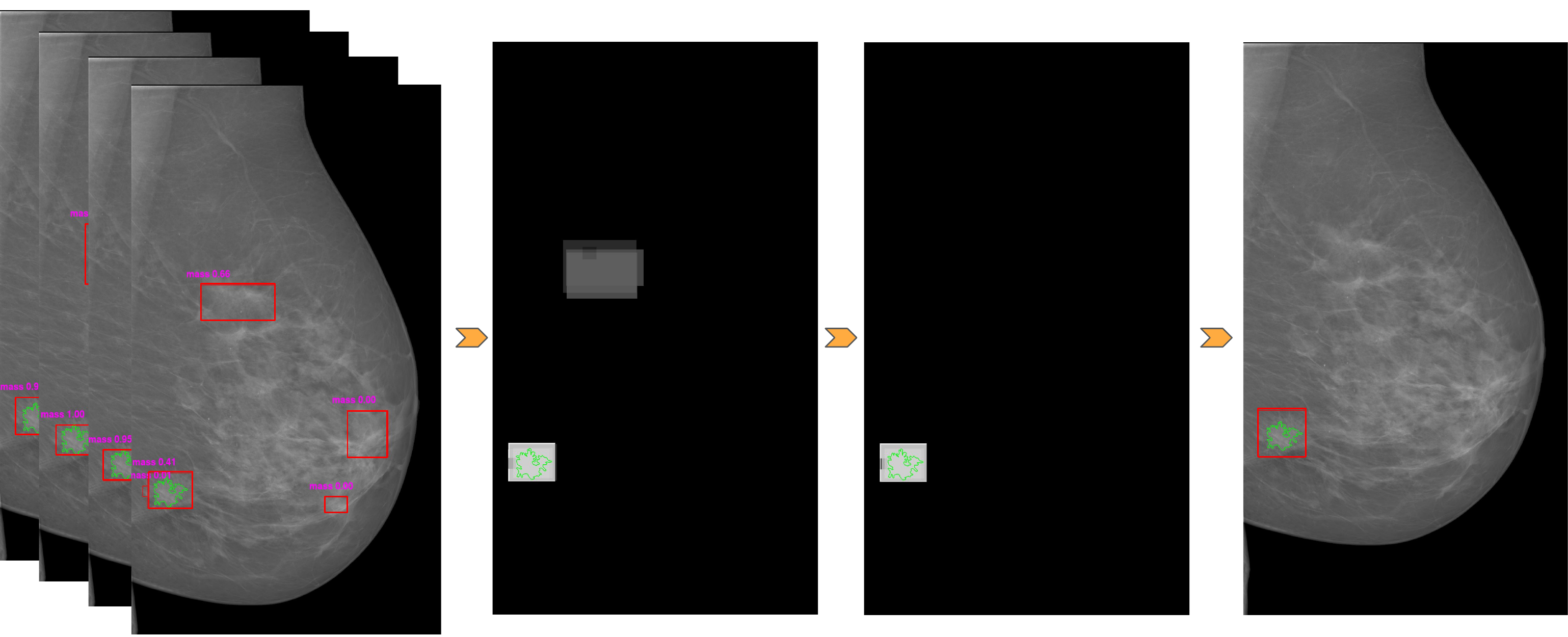, width = 12cm}} \vspace{-0.1cm} \\
\footnotesize \hspace{0.5cm} (a) \hspace{2.5cm} (b) \hspace{2.5cm} (c) \hspace{2.5cm} (d) \vspace{-0.1cm} \\
\caption{Proposed multi-scale fusion (MSF) applied to YOLOv3 predictions. The MSF strategy focuses on redundant information in multiple predictions. Red boxes correspond to mass RoI candidates. Green contours arise from ground truth annotations.}
\label{fig:Fig3}
\end{figure}

The proposed MSF scheme consists of three main steps (Fig.\ref{fig:Fig3}). For a given mammogram, detections are first carried out at different image scales (Fig.\ref{fig:Fig3}a). Since larger resolution will exceed the memory limits while smaller resolution will reduce the accuracy, we use the following 5 image ratios: $(160\times320)$, $(256\times512)$, $(320\times640)$, $(416\times832)$, $(480\times960)$. Second, we collect all $B$ coordinates of candidate bounding boxes and the corresponding confidence score sets $C$ provided in the previous step by YOLOv3. For each of these boxes $B_{i}$, we create a confidence mask $M_i$ where the value of the box region is the corresponding confidence score $ c_i$. Let $(X,Y)_i$ be the set of coordinates from bounding box $B_{i}$. For each $(x,y)$ in $(X,Y)_i$, we assign $M_i(x,y) = c_i$ with $c_i \in C$. After that, a single confidence mask $M_s$ (Fig.\ref{fig:Fig3}b) is created by fusing the set of confidence masks $ \{ M_1, M_2, \ldots, M_ {B} \}$ obtained at each prediction scale. $M_s$ is computed and normalized as follows:

\begin{equation}\label{eq:1}
M_s = \frac{\displaystyle \sum_{i=1}^{B} M_i}{N \times \max (c_1,c_2,\ldots,c_B)}
\end{equation} \vspace{-0.05cm} 

\noindent Third, we consider an empirically selected threshold $\lambda$ to implement majority voting (Fig.\ref{fig:Fig3}c) to the fusion mask $M_s$ by keeping areas where $M_s\geq\lambda$. Then, we measure the properties of labeled $M_s$ and find bounding box(es) that describe the fusion mask most properly (Fig.\ref{fig:Fig3}d), i.e. we find bounding box tuples ($\min_x, \min_y, \max_x, \max_y$) such that pixels of the same label belong to the same bounding box in the half-open intervals $[\min_x; \max_x)$ and $[\min_y; \max_y)$. 

Through the proposed MSF, we focus on redundant information that appears in multiple scales. From a statistical point of view, MSF allows to identify the most frequently detected regions in multiple predicted maps in order to limit false-positive predictions. Conversely, areas detected in few prediction maps or areas with low confidence scores are unlikely to be selected. Moreover, we analyze the effect of the empirical parameter $\lambda$ in order to keep an high level of sensitivity while improving specificity. Accordingly, we are able to remove most of the uncertainty and find the most reliable predictions. Final detections are resized to $256\times 256$ patches and fed into our second stage.

\subsection{Region-based mass segmentation}
\label{sec:sec3.3}

\noindent After the image-based mass detection stage, we propose a region-based mass segmentation stage that performs refined mass delineation from candidate patches using a deep convolutional encoder-decoder. Among recent advances of segmentation approaches, we implement a powerful deep architecture with nested and skip connections, following U-Net\texttt{++} \cite{zhou2018unet++}.

So far, general semantic segmentation in natural images has achieved great success \cite{Long_2015_CVPR,badrinarayanan2015segnet,zhao2017pyramid}. Recently proposed conditional residual U-Net \cite{li2018improved}, conditional GAN \cite{Singh2020} and cascaded U-Net \cite{yan2019embc} implemented for breast mass segmentation are all extensions of standard U-Net \cite{RFB15a}. Essentially, they share a key idea: shortcut connections from the encoder to the decoder that fuse downsampled features with upsampled features to recover high-level details more accurately. However, such models suffer from loss of space resolution details and semantic gap along skip connections. Rather than using standard shortcuts, the employed model builds connections through a series of nested dense convolutional blocks as a convolutional pyramid to enhance feature fusion. Concatenating intermediate subsequent layers bridges the semantic gap between feature maps. Then, a deep supervision is applied to prevent gradient vanishing issues in the middle part during back-propagation while ensuring a better segmentation accuracy.

The architecture is derived from standard U-Net: we employ in practice the vgg19 network as backbone for the encoder, which consists of 16 convolutional layers with repeated $3\times 3$ convolutions followed by ReLU activation function and $2\times 2$ max-pooling (3 fully-connected layers are not included). The decoder is symmetrically designed. The proposed mass segmentation method is referred as v19U-Net\texttt{++}. Since reaching a generic from-stratch model without overfitting is difficult, we pre-train the encoder branch using ImageNet \cite{deng2009imagenet} following \cite{conze2020healthy} to reduce the data scarcity issue while allowing faster convergence.  We exhaustively implemented four segmentation models for comparison: U-Net \cite{RFB15a}, cGAN \cite{Singh2020}, cascaded U-Net \cite{yan2019embc} as well as v19U-Net\texttt{++} as suggested. Once we get segmentation results, we can reconstruct high-resolution full mammograms with mass identification and delineation for visualization purposes.

\section{Experiments and results}
\label{sec:sec4}

\noindent In what follows, we report experimental settings and results for image-based detection (Sect.\ref{sec:sec4.1}) and segmentation (Sect.\ref{sec:sec4.2}) of breast masses. In particular, evaluations of final segmentation results are carried out both quantitatively and qualitatively. All experiments are implemented using Keras backend with a single Nvidia GeForce GTX 1080Ti GPU.

\subsection{Image-based mass detection}
\label{sec:sec4.1}

\noindent Experiments of this stage focus on mass detection from $2048\times1024$ mammograms. Typically, training a detection model on an insufficient dataset such as INbreast does not guarantee precise results. Therefore, a transfer learning technique is used to leverage a deep learning model on one task to another related task. In this work, we use convolutional weights pre-trained on ImageNet \cite{deng2009imagenet}, then we conduct transfer learning from DDSM-CBIS to INbreast. The DDSM-CBIS database is only employed in the detection stage, where all 1514 images containing masses are employed to pre-train the YOLOv3 model for 60k iterations before fine-tuning on INbreast for 30k iterations with batch size 32. The initial learning rate is set to 0.001 and decreases by 0.1 after 10k and 20k iterations. The INbreast dataset is too small to be representative if being divided into three subsets (train, validation and test sets). Therefore, we employ a ratio of 70\% to split INbreast into train and test subsets containing respectively 74 and 33 images. In order to eliminate the bias error, we use 5 random splits (denoted as T1, T2, ..., T5) to provide averaged results with cross-validation.\\

\begin{figure}[t]
\centering
\includegraphics[width=.95\textwidth,clip=true,trim = 0mm 0mm 0mm 13mm]{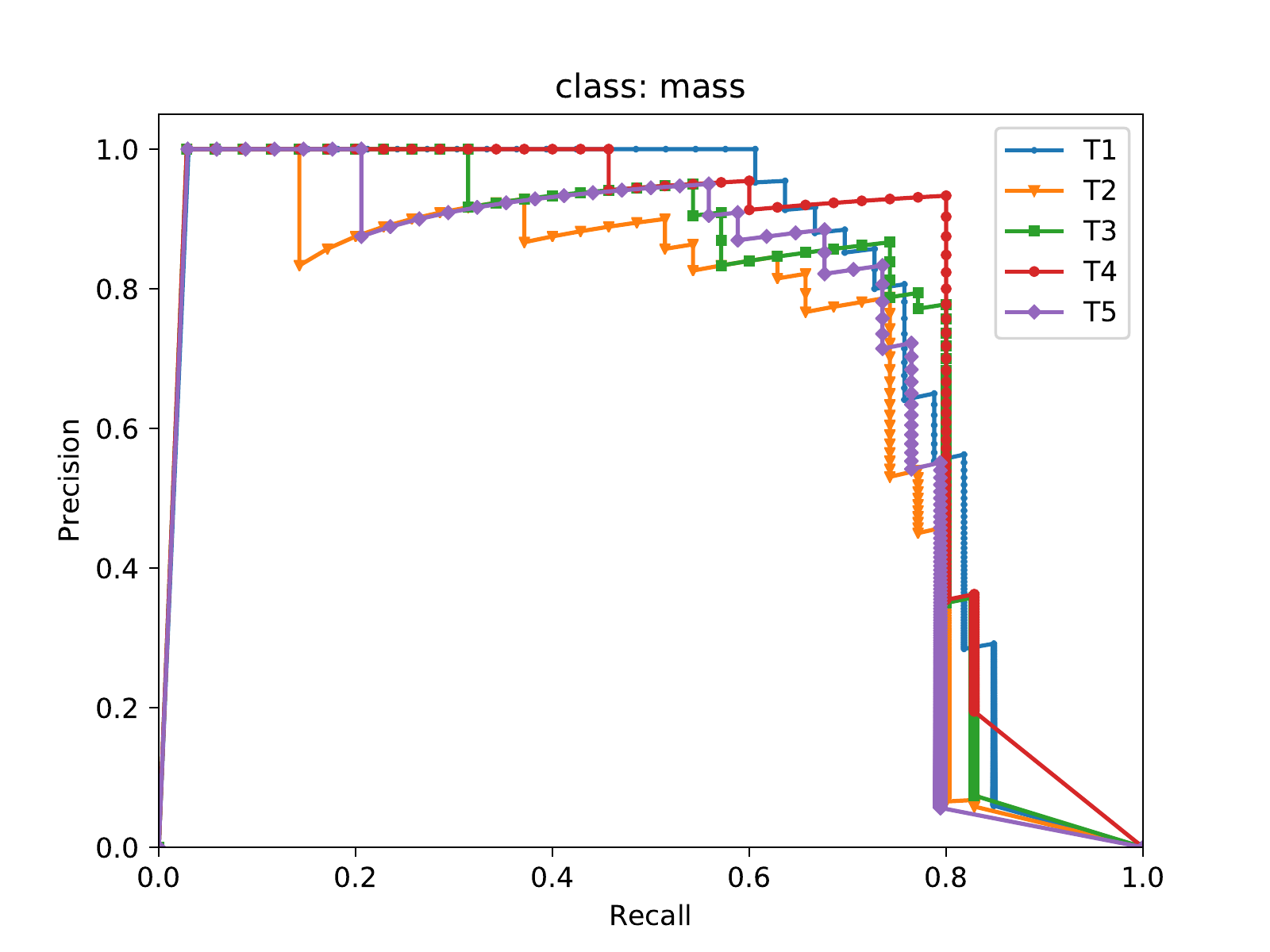} \vspace{-0.4cm}
\caption{Precision-recall curves of the YOLOv3 \cite{YOLOv3} detection results on 5 test sets (from T1 to T5) extracted from the INbreast \cite{Moreira2012INbreastTA} dataset.}
\label{fig:Fig4}
\end{figure}

\noindent \textbf{Mass localization evaluation.} We evaluate the detection performance of YOLOv3 by calculating the average precision (AP) score for masses present in each test set. Fig.\ref{fig:Fig4} shows precision-recall curves for each test set using an intersection over union $\geq$ 0.5. Precision-recall curves summarize the trade-off between the true positive rate and the positive predictive value using different probability thresholds. Then, we compute the average precision scores which summarize the weighted increase in precision with each change in recall for the thresholds in the precision-recall curve. From Fig.\ref{fig:Fig4}, we can clearly see that the precision-recall curves are fairly consistent between different test sets, which demonstrates the consistency of YOLOv3. Tab.\ref{tab:tab1} displays the corresponding AP scores of each curve. YOLOv3 yields an averaged AP of 75.46\% with a standard error of 1.7. For comparison, most state-of-the-art methods achieve a mean AP of 80\% on PASCAL VOC and 60\% on MS-COCO, which reveals very reasonable precision given the complexity of the mass detection task. 

\begin{table}
    \centering
    \begin{tabular}{ccccccc}
    \hline
    \textbf{Metrics} & T1 & T2 & T3 & T4 & T5 & average\\
    \hline
    AP (\%)  & 78.64 & 70.24 & 76.11 & 79.05 & 73.28 & 75.46$\pm$1.7\\
    \hline
    \end{tabular} \vspace{0.2cm}
    \caption{Performance of YOLOv3 \cite{YOLOv3} on the INbreast \cite{Moreira2012INbreastTA} dataset using average precision (AP) scores. T1 to T5 correspond to 5 experimental test sets.} \vspace{0.2cm}
    \label{tab:tab1}
\end{table}

\begin{figure}[t]
\centering
\includegraphics[width=.9\textwidth,clip=true,trim = 0mm 0mm 0mm 14mm]{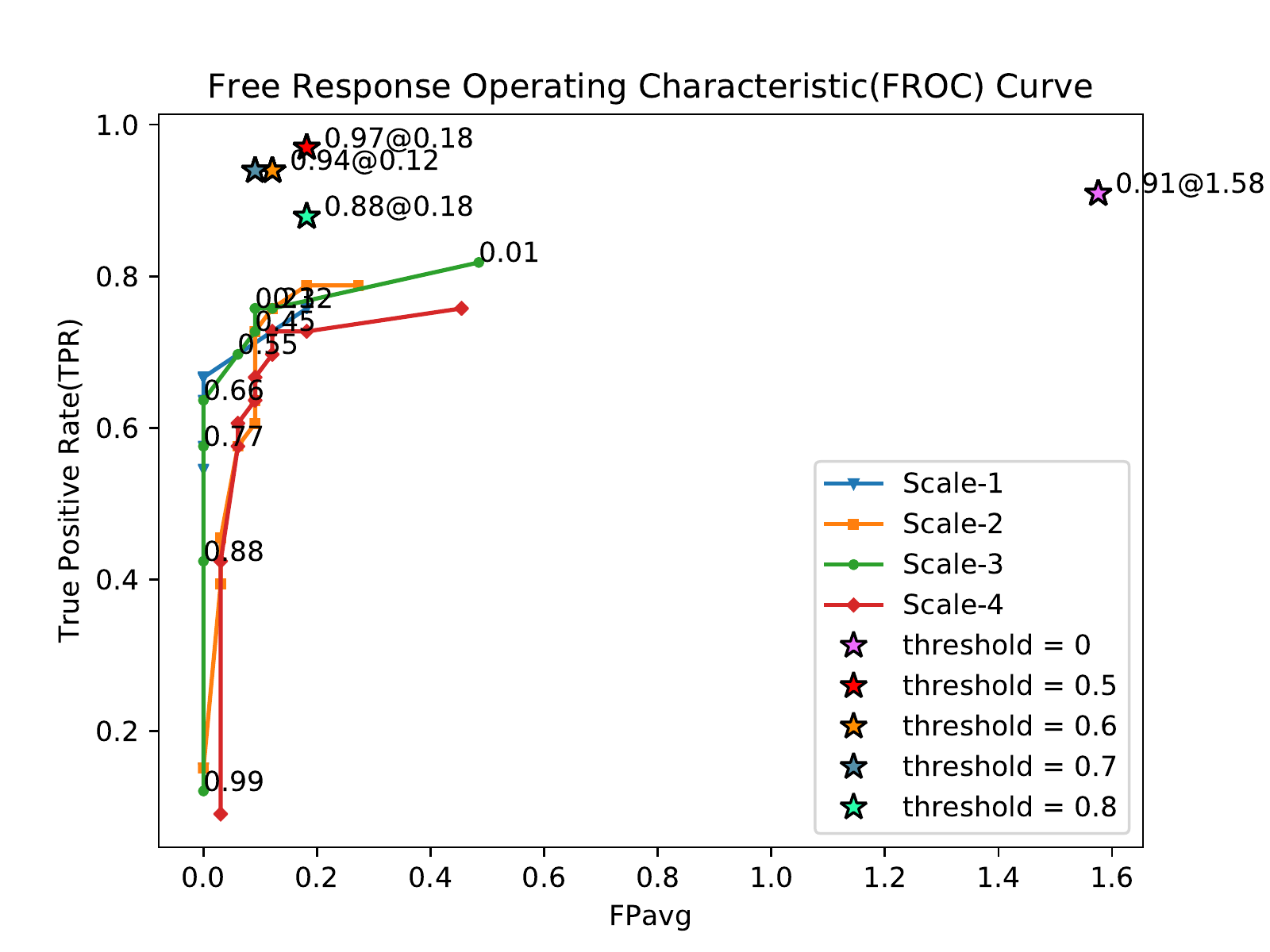} \vspace{-0.4cm}
\caption{Free response operating characteristic (FROC) curves of detection results on INbreast \cite{Moreira2012INbreastTA}, representing true positive rate (TPR) and average false positive per image (FPavg). Curves from Scale-1 to Scale-4 display results of single-scale predictions at $160\times320$, $256\times512$, $320\times640$ and $480\times960$. Stars shows TPR@FPavg of the final decision at fixed thresholds.}
\label{fig:Fig5}
\end{figure}

We fuse prediction results obtained at resolutions $160\times320$, $256\times512$, $320\times640$, $416\times832$ and $480\times960$ for multi-scale fusion (Sect.\ref{sec:sec3.2}). We use free-response receiver operating characteristic (FROC) as evaluation criterion.
Fig.\ref{fig:Fig5} illustrates the performance of MSF for test set T1 as example. The FROC curve is created by plotting the true positive rate (TPR) against the average false positive per image (FPavg) using various thresholds. Since MSF uses an empirical threshold $\lambda$ to make final decisions, we tested a set of thresholds $\lambda \in \{0,0.5,0.6,0.7\}$ to get different TPR@FPavg scores. $\lambda = 0$ means that we keep all the detections of YOLO, while $\lambda = 0.5$ means that we keep the part of mask $M_s\geq0.5$ (Eq.\ref{eq:1}) and so on. Fig.\ref{fig:Fig5} indicates that TPR@FPavg scores of MSF are all located in the upper left corner of FROC space, showing that our MSF strategy largely boosts the accuracy of mass localization compared to single-scale detections, with a more reliable TPR and less FP proposals. Additionally, the TPR@FPavg scores shown in Tab.\ref{tab:tab2} highlights the influence of $\lambda$. With a higher threshold, the false positives tend to be reduced while the TPR reaches the peak levels at around $\lambda = 0.5 \sim 0.6$. We finally choose $\lambda = 0.6$ considering the trade-off between true-positives and false-positives proposals.

\begin{table}
    \centering
    \begin{tabular}{lccccc}
    \hline
    \multirow{2}{*}{$\lambda$}&\multicolumn{5}{c}{\textbf{TPR@FPavg}}\\
    \cline{2-6}
     & T1 & T2 & T3 & T4 & T5\\
    \hline 
   \bm{$\lambda = 0$} & 0.91@1.58 & 0.97@1.39 & 0.89@1.30 & 0.91@1.55 & 1.0@0.87 \\
   \bm{$\lambda = 0.5$}& 0.97@0.18 & 0.94@0.27 & 0.91@0.18 & 0.92@0.36 & 0.97@0.12 \\ 
   \bm{$\lambda = 0.6$}& 0.94@0.12 & 0.94@0.24 & 0.91@0.18 & 0.92@0.30 & 0.97@0.06 \\
   \bm{$\lambda = 0.7$}& 0.94@0.09 & 0.89@0.27 & 0.91@0.15 & 0.89@0.18 & 0.97@0.06 \\
    \hline
    \end{tabular} \vspace{0.1cm}
        \caption{Performance of the proposed MSF method on INbreast \cite{Moreira2012INbreastTA} using TPR@FPavg scores with different $\lambda$. T1 to T5 correspond to the 5 experimental test sets.}
    \label{tab:tab2}
\end{table}

We also compare the image-based mass detection with respect to state-of-the-art using \texttt{TPR}@\texttt{FP}avg (Tab.\ref{tab:tab5}). Even if results are only for reference since datasets used for training and testing are not identical, it highlights that MSF ($0.94$@$0.22$) significantly outperforms \cite{sapate2020breast,ribli2018detecting,agarwal2019automatic} in both \texttt{TPR} and \texttt{FP}avg and shows consistent \texttt{TPR} with respect to \cite{Dhungel2017ADL} ($0.95$@$5$) while providing less FP. 

\begin{table}[t]
\centering
\begin{tabular}{llll}
\hline
\textbf{Methods}  & \texttt{TPR}@\texttt{FP}avg & dataset & images \\
\hline
Sapate et al. (2020) \cite{sapate2020breast} & 0.88 @1.51 & DDSM & 148 \\
\hline
Ribli et al. (2018) \cite{ribli2018detecting} & 0.90 @ 0.3 & INbreast & 107\\
\hline
Dhungel et al. (2017) \cite{Dhungel2017ADL} & 0.95 @ 5 & INbreast & 410\\
\hline
Agarwal et al. (2019) \cite{agarwal2019automatic} & 0.92 @ 0.5 & INbreast & 410 \\
\hline
YOLOv3+MSF (ours) $\lambda=0.5$ & 0.94 @ 0.22  & INbreast & 107 \\
\hline
\end{tabular}
\caption{Detection performance comparisons between the proposed MSF and state-of-the-art \cite{sapate2020breast,ribli2018detecting,Dhungel2017ADL,agarwal2019automatic}. Our provided \texttt{TPR}@\texttt{FP}avg score is the average of T1 to T5 test sets at $\lambda = 0.5$. }
\label{tab:tab5}
\end{table}

\subsection{Mass segmentation}
\label{sec:sec4.2}

\noindent \textbf{Region-based mass segmentation.} We perform extensive experiments on INbreast \cite{Moreira2012INbreastTA} to validate the employed CED network with nested and dense skip connections (v19U-Net\texttt{++}, Sect.\ref{sec:sec3.3}). We compared it with the baseline U-Net \cite{RFB15a} as well as two other recently published architectures: cGAN \cite{Singh2020} and cascaded U-Net \cite{yan2019embc}. Experiments are carried out using the same train-test splits on INbreast examinations as in the previous stage. Training image crops are extracted around ground truth masses and resized to $256\times256$ pixels. Histogram equalization is then used to enhance the contrast. We train each model with a batch size of $4$, Adam optimizer and Dice loss (the cGAN network loss is formulated by combining binary cross entropy and Dice losses) defined as $\frac{2TP}{2TP+FP+FN}$ where TP, FP, TN, and FN are the pixel level true positives, false positives, true negatives and false negatives. We use pre-trained weights from ImageNet \cite{deng2009imagenet} and then train models until convergence.

To assess the final segmentation performance, we compute Dice scores over each test set on full mammograms for each different methodology (Tab.\ref{tab:tab3}). Compared to U-Net \cite{RFB15a} (89.20$\pm$0.5), results of cascaded U-Net (89.49$\pm$0.3) are slightly better since it employs a multi-scale cascade of U-Net combing auto-context \cite{yan2019embc}. The gain is relatively low considering that \cite{yan2019embc} has been designed to tackle mass segmentation from entire mammograms. cGAN \cite{Singh2020} also brings slight benefits (90.02$\pm$0.2) to the original U-Net but less than v19U-Net\texttt{++} \cite{zhou2018unet++} which yields the best results on all test sets with 90.86\% as average Dice score. \\

\begin{table}
    \centering
    \begin{tabular}{lcccccc}
    \hline
    \textbf{Methods} & T1 & T2 & T3 & T4 & T5 & average (\%)\\
    \hline
    U-Net \cite{RFB15a} & 90.47 & 89.76 & 88.16 & 87.97 & 89.66 & 89.20$\pm$0.5\\
    cGAN \cite{Singh2020} & 90.30 & 90.53 & 89.70 & 89.33 & 90.22 & 90.02$\pm$0.2\\
    cascaded U-Net \cite{yan2019embc} & 89.20 & 90.40 & 88.83 & 89.18 & 89.82 & 89.49$\pm$0.3\\
    v19U-Net\texttt{++} \cite{zhou2018unet++} & \textbf{90.94} & \textbf{91.42} & \textbf{90.56} & \textbf{90.23 }& \textbf{91.13} & \textbf{90.86$\pm$0.2}\\
    \hline
    \end{tabular} \vspace{-0.15cm} 
    \caption{Average Dice score (\%) of different patch-based deep segmentation methods on INbreast \cite{Moreira2012INbreastTA} mass patches centered around ground truth masses. Best scores are in bold.}
    \label{tab:tab3}
\end{table}

\begin{table}[t]
\hspace{-1.5cm} \small{
\begin{tabular}{l|l|ccccc|c} 
\hline
\textbf{Method} & \textbf{Setup} & T1 & T2 & T3 & T4 & T5 & average (\%)\\
\hline
\multirow{3}{*}{U-Net \cite{RFB15a}} & one-stage & 43.66 & 44.12 & 45.93 & 40.79 & 47.36 & 44.37$\pm$1.1\\
& two-stage w/o MSF & 70.59 & 68.46 & 70.56 & 74.66 & 66.06& 70.07$\pm$2.8\\
& two-stage with MSF & \textbf{77.40} & \textbf{83.07} & \textbf{75.45} & \textbf{77.80} & \textbf{82.47} & \textbf{79.24$\pm$1.5}\\
\hline
\multirow{3}{*}{cGAN \cite{Singh2020}} & one-stage & 25.27 & 30.91 & 24.74 & 23.21 & 40.45& 28.92$\pm$3.2\\
& two-stage w/o MSF &70.28 &	66.93 & 70.22 & 74.93 & 63.73 &69.22$\pm$3.7\\
& two-stage with MSF &	\textbf{75.66} &\textbf{81.66} & \textbf{76.70} & \textbf{77.44} & \textbf{83.45} &  \textbf{78.98$\pm$1.5}\\
\hline
\multirow{3}{*}{cascaded U-Net \cite{yan2019embc}} & one-stage &64.37 & 61.56 & 65.63 & 65.35 & 70.55 & 65.49$\pm$1.5\\
& two-stage w/o MSF & 70.89 & 67.78 & 70.01 & 73.35 & 65.02 &69.81$\pm$3.4\\
& two-stage with MSF &  \textbf{75.76} & \textbf{82.51} & \textbf{76.78}  & \textbf{77.69} & \textbf{83.16} &\textbf{79.18$\pm$1.5}\\
\hline
\multirow{3}{*}{v19U-Net\texttt{++} \cite{zhou2018unet++}}& one-stage & 53.38 & 49.38 & 47.44 & 48.85 & 61.80 & 52.17$\pm$2.6\\
& two-stage w/o MSF & 72.18 & 68.55 & 72.27 &76.10 & 65.69 & 70.96$\pm$3.6 \\
& two-stage with MSF & \textbf{77.51} & \textbf{84.38} & \textbf{77.39} &	\textbf{78.80} & \textbf{84.12}& \textbf{80.44$\pm$1.6} \\
\hline
\end{tabular}} \vspace{-0.1cm}
\caption{Average Dice score (\%) obtained on final delineations from 2048$\times$1024 full INbreast \cite{Moreira2012INbreastTA} mammograms. Best scores are in bold.}
\label{tab:tab4}
\end{table}
 
\noindent \textbf{Two-stage mass segmentation.} To assess the final segmentation performance of the proposed two-stage system (Fig.\ref{fig:Fig1}), we compare the overall Dice on full mammograms from different methods. As a proof of concept, we test the second stage (Sect.\ref{sec:sec3.3}) using the candidate patches arising from the first stage (Sect.\ref{sec:sec3.2}), which are resized to $256\times256$ pixels before feeding into segmentation models. Tab.\ref{tab:tab4} presents comparative evaluations for each model: one-stage segmentation, two-stage segmentation without and with the proposed MSF on high-resolution full mammograms. In particular, in the two-stage without MSF setup, mass candidates are provided by a simple single-scale prediction of YOLOv3.

\begin{figure}[t]
\centering
\subfigure{
\centering
\begin{minipage}[t]{0.03\linewidth}
\vspace{-3cm} \rotatebox{90}{\small (a)}
\end{minipage}
\includegraphics[width=0.9\textwidth]{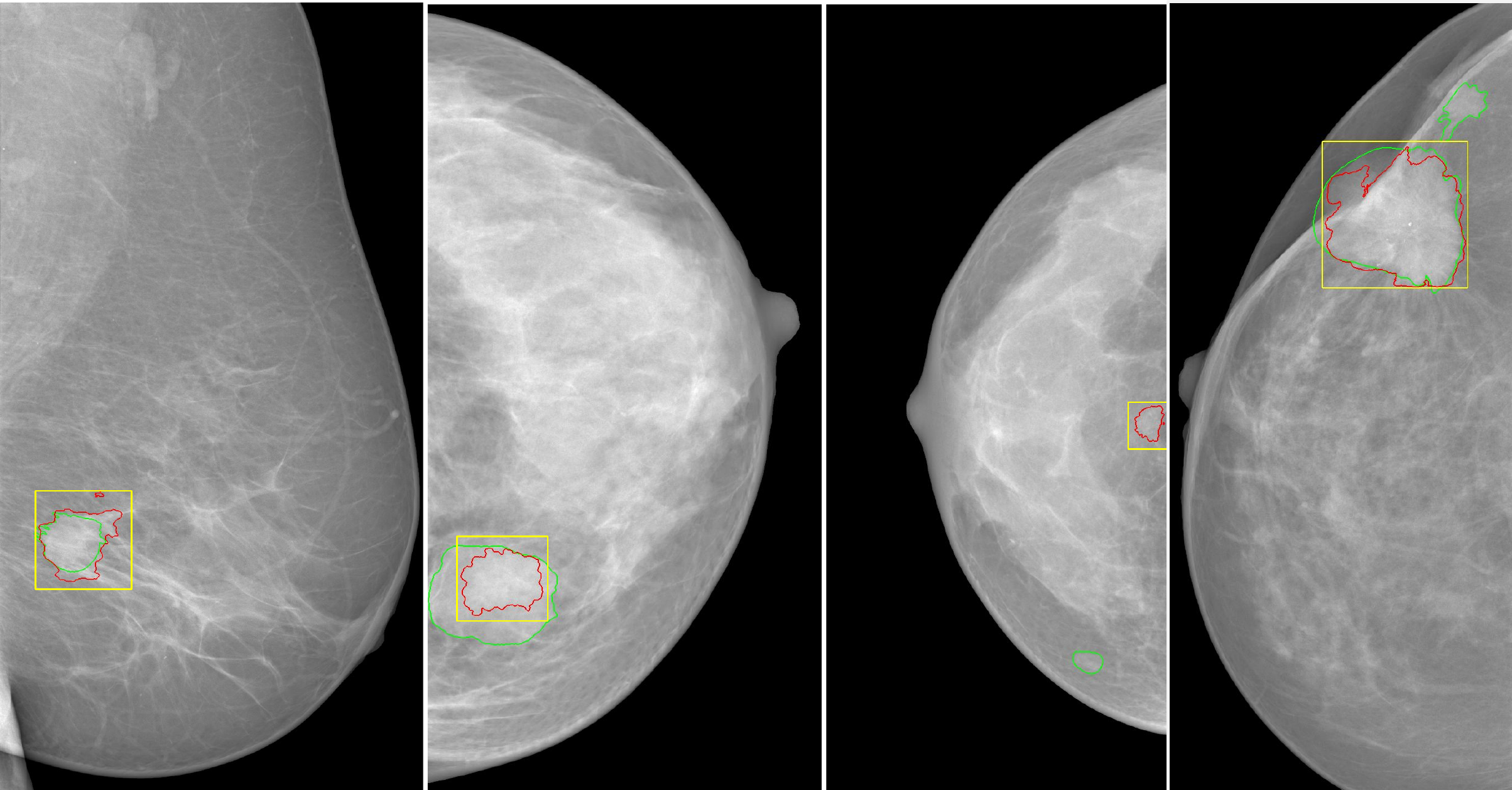}}
\vspace{-0.2cm} \\
\subfigure{
\centering
\begin{minipage}[t]{0.03\linewidth}
\vspace{-3cm} \rotatebox{90}{\small (b)}
\end{minipage}
\includegraphics[width=0.9\textwidth]{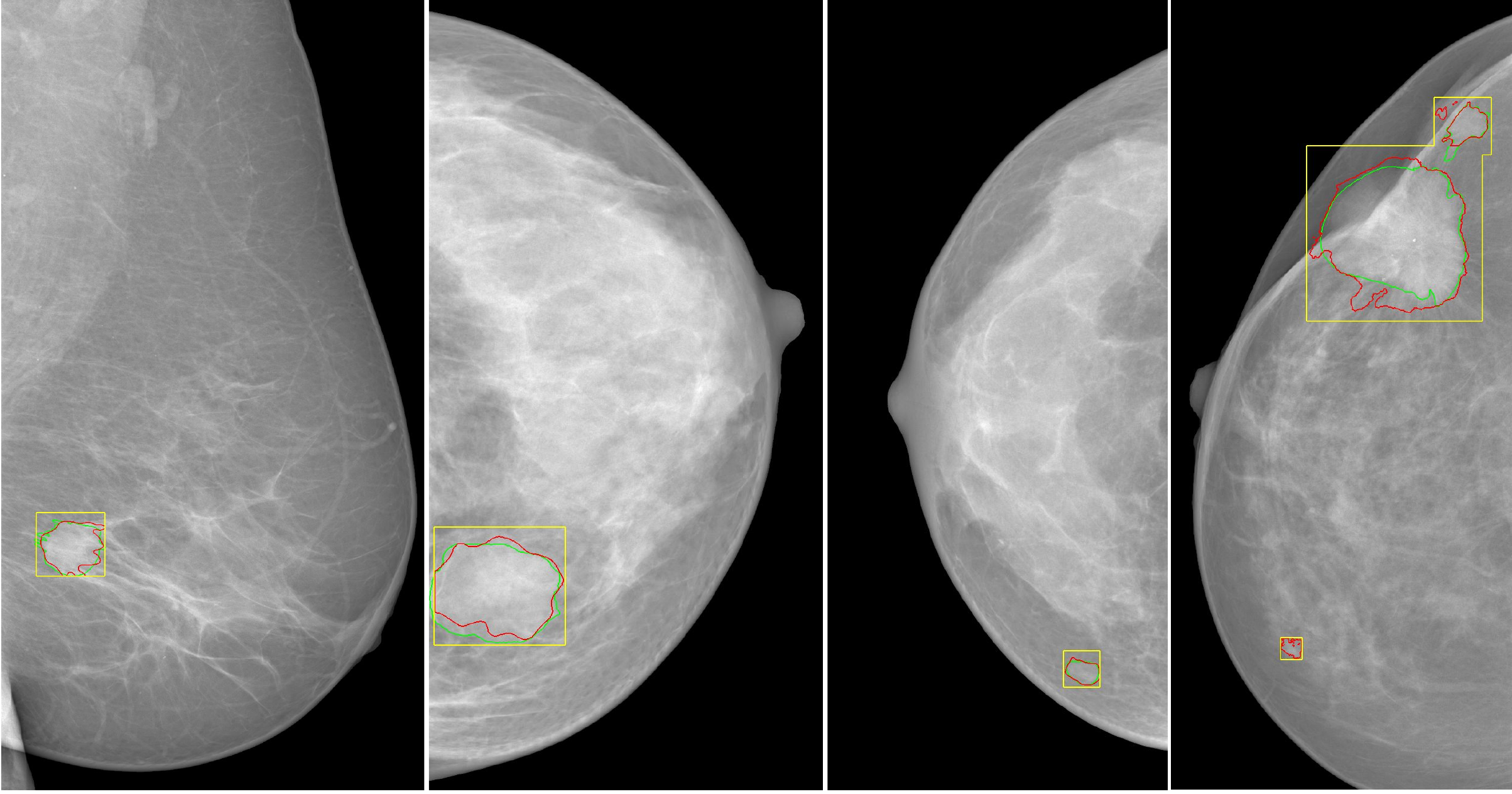}} \vspace{-0.2cm}
\caption{Mass segmentation using our two-stage method without (a) and with (b) multi-scale fusion (MSF). Yellow, red and green stand for final detection, segmentation and ground truth.}
\label{fig:Fig6}
\end{figure}

\begin{figure}[t]
\centering
\subfigure{
\centering
\begin{minipage}[t]{0.03\linewidth}
\vspace{-3cm}
\rotatebox{90}{\small (a)}
\end{minipage}
\includegraphics[width=0.9\textwidth]{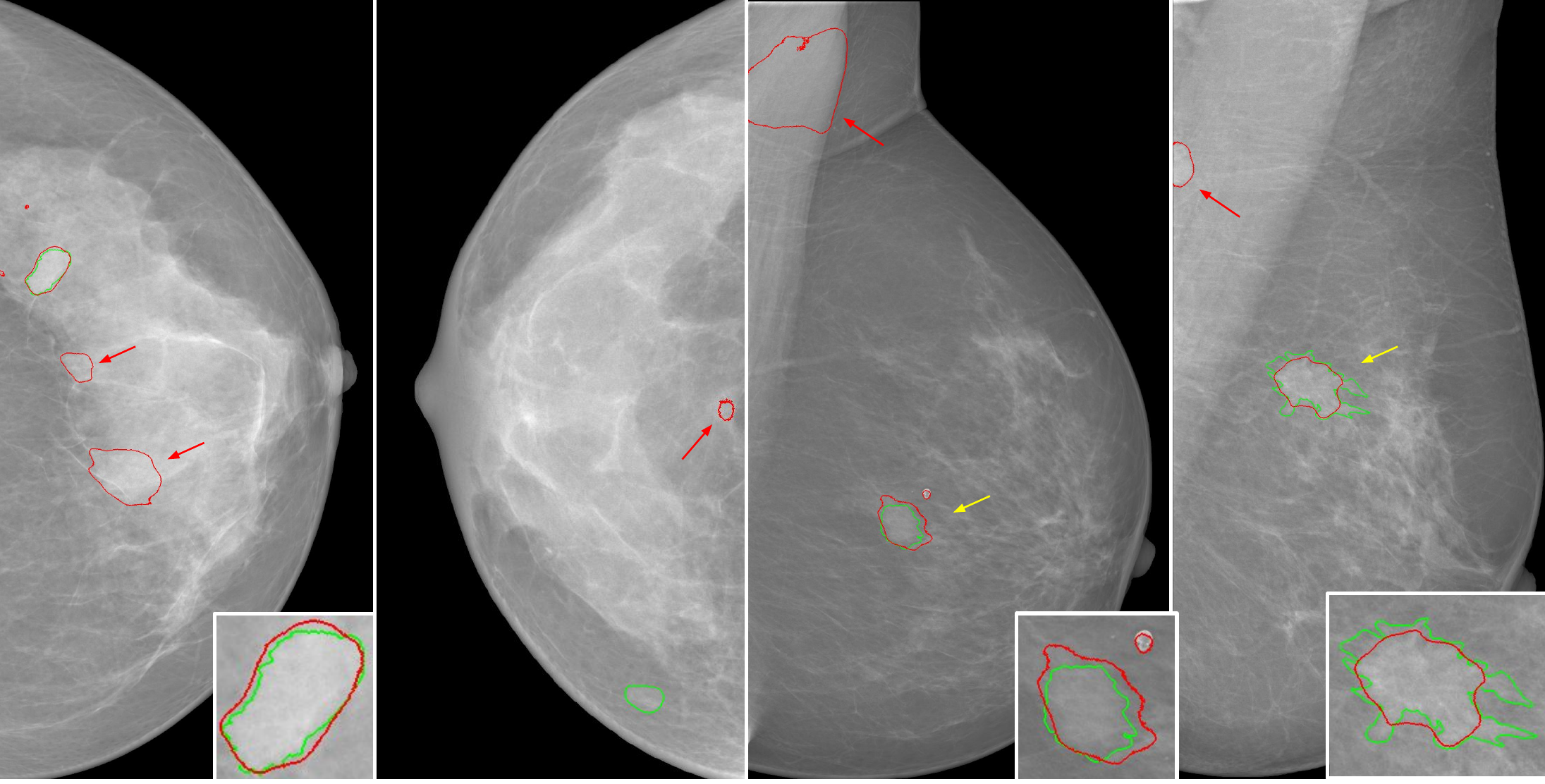}} \vspace{-0.2cm} \\
\subfigure{
\centering
\begin{minipage}[t]{0.03\linewidth}
\vspace{-3cm}
\rotatebox{90}{\small (b)}
\end{minipage}
\includegraphics[width=0.9\textwidth]{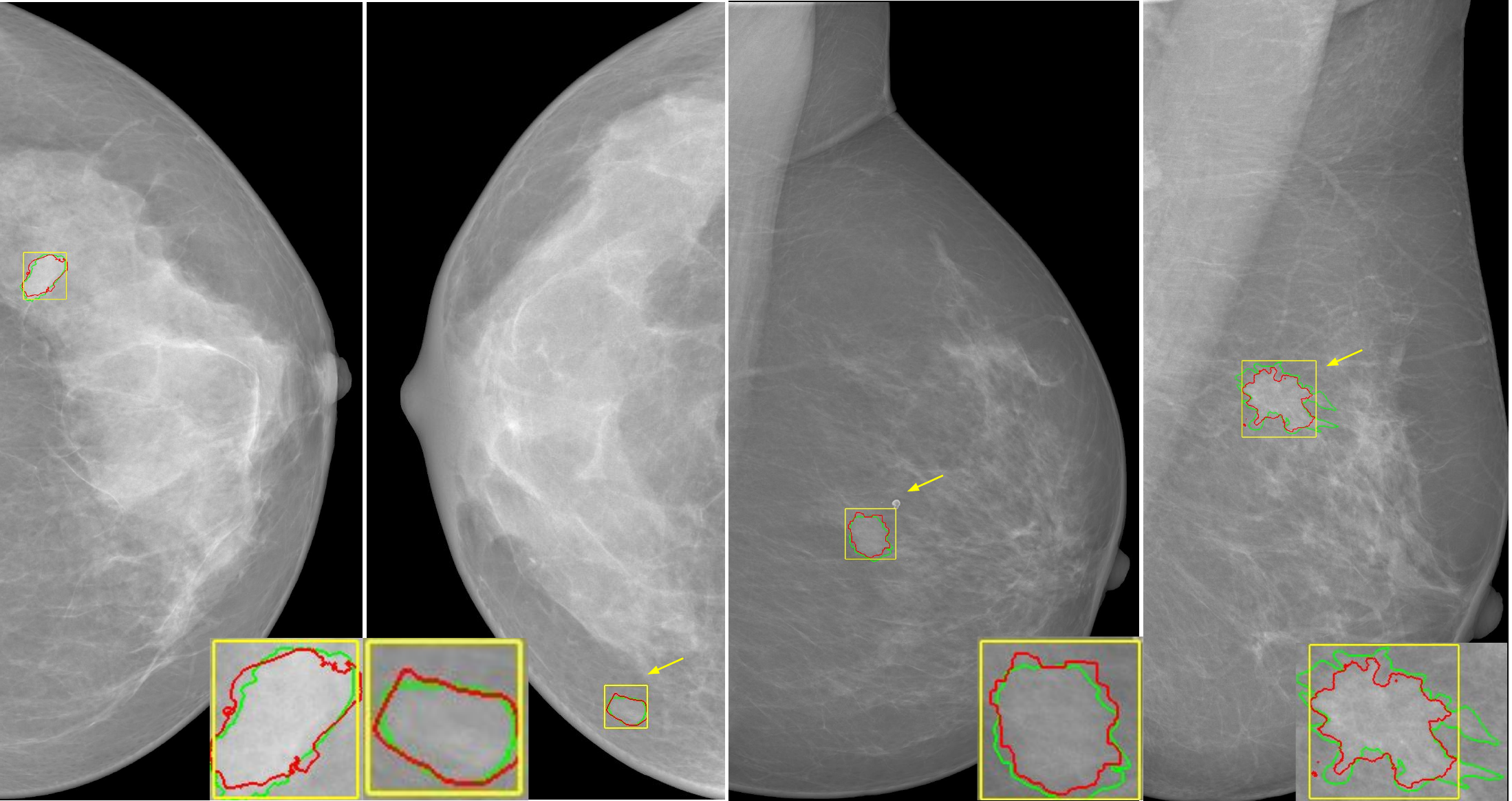}} \vspace{-0.15cm}
\caption{Mass segmentation using cascaded U-Net \cite{yan2019embc} (a) and our two-stage method with MSF (b) on INbreast \cite{Moreira2012INbreastTA} images. Yellow, red and green lines stand for final detection, segmentation and ground truth contours. Yellow (red) arrows highlight true-positive (false-positive) cases.}
\label{fig:Fig7}
\end{figure}

Comparisons between models indicate that v19U-Net\texttt{++} yields better segmentation results for two-stage segmentation, with an average Dice score of 70.96\% without MSF and 80.44\% with MSF. Compared with one-stage segmentation, a significant gap is crossed when using a two-stage scheme, demonstrating the effectiveness of our two-stage localization-segmentation design. MSF brings Dice improvements to the two-stage scheme from 9.17\% with U-Net to 9.76\% with cGAN (9.48\% with v19U-Net\texttt{++}), showing that adding the MSF strategy into the pipeline can further greatly improve performance. We also observe that one-stage segmentation methods reach various level of robustness \cite{yan2019embc} when applied to high-resolution mammograms: from 28.92\% (cGAN) to 65.49\% (cascaded U-Net). Conversely, our two-stage scheme provides more stable and reliable results, which suggests that it could be very effective in clinical practice. 

Evaluation is supplemented with qualitative results. Fig.\ref{fig:Fig6} shows full mammogram detection and segmentation results using the proposed two-stage with MSF compared to two-stage without MSF. We observe that by using the MSF strategy, we have considerable improvements in both mass localization accuracy and mass delineation precision. It also shows that we can successfully detect multiple masses in a single mammogram. In addition, we compare in Fig.\ref{fig:Fig7} the proposed method with cascaded U-Net \cite{yan2019embc} which also addresses full mammogram segmentation. Our method obtains more accurate detections and boundary adherence, while almost all false-positive proposals are eliminated. Moreover, the method is robust in dealing with masses of any size, shape or texture. This confirms that our methodology is very generalizable in handling the problem of strong class imbalance and tumor appearance variability. 

\section{Discussion}
\label{sec:sec5}

\noindent When dealing with breast mass segmentation from full mammograms at native resolution, one-stage segmentation appears impractical due to the contradiction between the preservation of high-level semantic information and resolution details. In turn, CAD systems integrated into routine requires high accuracy due to clinical requirements, i.e. high true positive rate combined with low false positive rate. Meanwhile, the feasibility is also a key aspect that should not be overlooked towards efficient deployment. The ideal CAD system should be able to help with diagnosis without any additional radiologist guidance. 

In this work, we came up the idea of a two-stage method which is desired to imitate the realistic procedure in clinical scenarios, and we tried to automatize the candidate selection process using multi-scale fusion. First, the deep network roughly localizes masses of any size, position and shape from the whole image by fusing predictions at multiple scales. Second, we perform an effective patch-based deep segmentation method with nested and dense shortcuts to obtain the accurate delineation of mass contours. Our system is able to achieve 80.44\% Dice, which sets the state-of-the-art performance in mass segmentation on the publicly available INbreast dataset, outperforming one-stage segmentation schemes such as cGAN \cite{Singh2020} (28.92\%), U-Net \cite{RFB15a} (44.37\%) or cascaded U-Net \cite{yan2019embc}) (65.49\%). The newly designed MSF brings Dice improvements to the two-stage scheme from 9.17\% (U-Net\cite{RFB15a}) to 9.76\% (cGAN \cite{Singh2020}). Fusing predictions performed at multiple scales addresses the problem of unsuccessful single detection and avoids manual selection (contrary to \cite{Alantari2018AFI}) by removing the most of the uncertainty while finding the most reliable abnormalities without interventions. This makes our method very competitive for integration into clinical practice.

\section{Conclusion}
\label{sec:sec6}

\noindent In this paper, we studied the problem of automated mass segmentation from high-resolution full mammograms. We proposed a two-stage framework combining a deep, coarse-scale mass detection with a new multi-scale fusion strategy and a fine-scale mass segmentation using dense and nested skip connections. Our system works as an accurate and automatic mass localization and segmentation CAD system. Results on the INbreast dataset confirm that the proposed pipeline outperforms state-of-the-art with promising model robustness and generalizability. Our contributions make full-mammogram mass segmentation more reliable and steadily push forward the implementation of realistic CAD systems.

Future research should consider the potential effects of fusing multi-view and contralateral symmetry information to increase the robustness of breast lesion detection and delineation and therefore improve clinical guidance.  Furthermore, our framework is generic enough to be extended to other medical imaging modalities for both anatomical and pathological structure segmentation.

\section*{Compliance with ethical standards} 

\noindent This research study was conducted retrospectively using human subject data made available in open access \cite{Moreira2012INbreastTA,ddsm-cbis}. None of the authors of this manuscript have any financial or personal relationships with other people or organizations that could inappropriately influence and bias this work. 

\bibliography{yan-BBE-2020}

\end{document}